\documentstyle[preprint,eqsecnum,aps,psfig]{revtex}

\begin{document}
\draft
\title{Scaling theory of Tomonaga-Luttinger liquid with 
$1/r^\beta$ type long-range interactions}
\author{Hitoshi Inoue}
\address{
Department of Physics, Kyushu University 33, Hakozaki,
Higashi-ku, Fukuoka 812-8581, Japan
}
\date{\today}
\maketitle
\begin{abstract}
We discuss effects of $1/r^\beta$ type  
long-range (LR) interactions in a tight-binding model by utilizing 
the bosonization technique,
renormalization group and conformal field theory (CFT). We obtain the 
low energy action known for Kibble's model which
generates the mass gap in 3 dimension 
when $\beta =1$, the Coulomb force case. In one dimension, 
the dispersion relations predict that the 
system remains gapless even for $\beta =1$ and 
the existences of Tomonaga-Luttinger liquid (TLL) when $\beta > 1$. 
When $\beta =1$,  
the LR interactions break TLL in the long 
wavelength limits, even if the strength 
is very small. We make the more precise 
arguments from the stand point of the renormalization group and CFT. 
Finally we derive the accurate finite size scaling of energies and  
thermodynamics. Moreover 
we proceed to numerical calculations, considering 
the LR umklapp process terms. We conclude that the TLL phase 
become wider in the strength space of 
interactions as the power $\beta$ approaches to $1$. 
\end{abstract}
\narrowtext
\section{Introduction}
The electron systems have attracted our much attention 
in the low and high energy physics.
As the dimension of the electron systems decrease,
the charge screening effects become less important. 
In spite of these facts 
models with short range interactions 
have been adopted in many researches 
of one dimensional electron systems.  
The recent advanced technology make it possible 
to fabricate quasi-one-dimension systems. Actually 
in a low temperature
the effect of Coulomb forces have been observed  
in $GaAs$ quantum wires \cite{GaAs},
quasi-one-dimensional conductors \cite{1Dcond1,1Dcond2,1Dcond3}
and 1D Carbon nanotubes \cite{ctube1,ctube2,ctube3}. 

A role of a $1/r$ Coulomb repulsive force
was investigated on long distance properties by bosonization techniques
\cite{Schulz}. The charge correlation 
decays very slowly with distance suggesting that the 
ground state is the {\it Wigner crystal} rather than 
the Tomonaga-Luttinger liquid (TLL). 

On one hand the 
insulator-metal transition caused by the long-range (LR) Coulomb
interactions have been discussed \cite{Maekawa}. For $1/r^2$ 
type weak interactions the ground state is TLL which is explained
by the Gaussian CFT \cite{Kawakami,Sutherland}.
It was reported that the 
strong $1/r^2$ interactions make the system 
gapless to gapful through the generalized Kosterlitz-Thouless 
transition \cite{Hatsugai}. 

The aim of this paper is to find the precise finite size  
scaling of energies and estimate 
the range of TLL in the strength space of 
the $1/r^\beta (\beta \geq 1)$ type
LR interactions for various powers $\beta$. 
The strategy is as follows. 
In section ${\rm I\!I}$ A and B
we discuss the long wavelength 
behaviors of the system when we consider the LR
forward scatterings by making use of bosonization techniques and the
renormalization group (RG) method. In the
section ${\rm I\!I}$ C        
we find the corrections caused by such the $\beta >1$ LR forwards scatterings
in the finite size 
scaling of energies and the thermodynamics.
In the section ${\rm I\!I\!I}$, 
we analyze a tight-binding model with the $\beta >1$ type 
LR interactions numerically 
by considering these corrections 
in the finite size scaling. 
We show numerical data of the drude weight and 
compressibility and determine the range of TLL in the interactions
strength space.
In section ${\rm I\!V}$ we argue the case $\beta =1$.
\section{FIELD THEORETICAL APPROACHES}
\subsection{Low energy action}
Schulz analyzed the effect of LR
Coulomb interactions by bosonization 
technique \cite{Schulz}. 
We extend the action to general $1/r^{\beta}$ type
interactions. Namely we write the action:
\begin{eqnarray}
 S &=& \int d\tau dx \frac{1}{2 \pi K} (\nabla \phi)^2 + g\int d\tau dx dx^{'}
\partial_{x} \phi(x,\tau)V(|x-x^{'}|)
\partial_{x^{'}} \phi(x^{'},\tau), \nonumber \\
\end{eqnarray}
where $V(x)=\frac{1}{|x|^\beta}$ and
$K$ is the TL parameter. The second term in (2.1) comes from the 
forward scattering in the fermion picture. 
Precisely speaking, there are 
other oscillating terms in the effective action. Now we focus on what
happens in considering only the forward scattering terms
of the charge density freedom. 
We shall discuss
the effects of the oscillating terms later in the section ${\rm I\!I\!I}$.
This action is known for Kibble's model \cite{Kibble},
the interactions of which
induce the mass gap in three dimension when $\beta =1$. 
The situations in one dimension 
are different from that in three dimension. 

To discuss in the Fourier space 
we choose the form 
$V(x)=\frac{1}{(x^2+\alpha^2)^{\beta /2}}$ which contains the cut-off
$\alpha$ in order to remove the ultra-violet divergences.
The expression in the wave number space of this action is
\begin{eqnarray}
S &=& \int dq dw \{ \frac{2 \pi}{K}(q^2+w^2)+ g q^2 V(q) \}|\phi(q,w)|^2,
\end{eqnarray}
where $V(q)$ is the Fourier transformation of $V(x)$ in one dimension.
From this the dispersion relation is 
\begin{eqnarray}
w^2 &=& q^2\{ 1+\frac{g K }{2 \pi}V(q)\}. 
\end{eqnarray}
We can derive the long wavelength behavior of $V(q)$:
\begin{eqnarray}
 V(q) &\sim& -A+B \ln q\;\;{\rm for \; \beta =1}\nonumber \\
      &\sim& C+D q^{\beta-1}\;\; {\rm for \; \beta > 1
(\beta \neq odd\;\;integer)}, 
\nonumber \\
      &\sim& E+ F q^{2} \ln q +G q^{2}+\cdots \;\;{\rm for \; \beta =3}
\nonumber \\
      &\sim& I+ J q^{2}+ K q^{4}\ln q+ Lq^{4}+\cdots 
\;\;{\rm for \; \beta =5} \nonumber \\ 
&& \cdots,		      	
\end{eqnarray}
where $A,\cdots ,L$ are constant (See APPENDIX A). 
From this we see that the system is 
gapless when $\beta \geq 1$ and it is expected to be 
TLL when $\beta > 1$, that is, $w \sim q$. 
When $\beta =1$, the LR interactions drive  
the ground state from TLL to the {\it Wigner crystal}\cite{Schulz}.
The slowest decaying part of the density 
correlation function is given by 
\begin{eqnarray}
 <\rho(x) \rho(0) > &\sim& \cos (2k_{F} x) {\exp (-{\rm c} 
{\sqrt {\log x}})},
\end{eqnarray}
where $c$ is a function of K \cite{Schulz}.
This feature is not like that of TLL:
\begin{eqnarray}
 <\rho(x) \rho(0) > &\sim& -K \frac{1}{x^2}+ 
{\rm const.} \frac{\cos (2k_{F} x)}{|x|^K}. 
\end{eqnarray}
\subsection{The treatments by the  renormalization group}
For the moment we focus our argument on the case $\beta > 1$ 
and $\beta \neq$ odd integer which 
is expected to belongs to
TLL from the dispersion relations (2.4). 
As the C term in similarities (2.4) give 
the linear dispersion, the C term is marginal. 
To see the essence of LR interactions
we separate the interactions term $g$ into two parts: 
\begin{eqnarray}
V(q) &=& \int dx \frac{e^{iqx}-1}{(x^2+\alpha^2)^{\beta/2}}
+ \int dx \frac{1}{(x^2+\alpha^2)^{\beta/2}} \nonumber \\
 &\equiv& V_{\rm long}(q)+V_{\rm short}.
\end{eqnarray}
Equivalently the expression in real space is 
\begin{eqnarray}
V(x) &=& V(x)-V_{\rm short}\delta(x)+V_{\rm short}\delta(x) \nonumber \\
&\equiv& V_{\rm long}(x)+V_{\rm short}\delta(x).
\end{eqnarray}
Hence the present action is rewritten to
\begin{eqnarray}
 S &=& \int d\tau dx \frac{1}{2 \pi K^{'}} 
(\nabla \phi)^2 + g\int d\tau dx dx^{'}
\partial_{x} \phi(x,\tau)V_{long}(x-x^{'})\partial_{x^{'}} \phi(x^{'},\tau). 
\nonumber \\
   &=&  \int dq dw \{ \frac{2 \pi}{K^{'}}(q^2+w^2)+ g q^2 
V_{long}(q) \}|\phi(q,w)|^2.
\end{eqnarray}
We derive the RG eq. of g:
\begin{eqnarray}
\frac{dg}{dl} &=& (1-\beta) g
\end{eqnarray}
(See APPENDIX B.).
Just later we also find the consistency of this eq. 
by conformal field theory.

For the case $\beta = 1$ we can derive the RG eq. of $g$ and the
velocity(See APPENDIX B.).
\subsection{The finite size scaling of energies}
We know the energy size scaling in the Gaussian CFT 
\cite{Cardy1,Cardy2,Cardy3,Cardy4}:
\begin{eqnarray}
   \Delta E_{n} &=& \frac{2\pi v x_{n}}{L} \nonumber \\
    E_{g} &=& e_{g}L -\frac{\pi v c}{6L}.
\end{eqnarray} 
Considering 
the LR interactions,
we can extract the corrections from these 
energy size scalings precisely (See APPENDIX C.): 
\begin{eqnarray}
\Delta E_{n} &=& \frac{2\pi v x_{n}}{L}(1+g\frac{{\rm const.}}{L^{\beta-1}}
+g \frac{1}{L^2}+O(1/L^2))
\nonumber \\
E_{g} &=& e_{g}L -\frac{\pi v c}{6L}(1+g\frac{{\rm const.}}{L^{\beta-1}}
+g \frac{1}{L^2}+O(1/L^2)),
\end{eqnarray}
where $\beta (> 1)$ is not odd integer. For the $\beta=$ odd integer, the 
logarithmic corrections appear. The details are shown in APPENDIX C
and the $\beta =1$ case is discussed later.
The $O(1/L^2)$ terms come from the irrelevant field
$L_{-2} \bar{L}_{-2} {\bf 1}$ and the LR g term. 
The first scaling of eqs. (2.12) implys that the LR forward scatterings
\begin{eqnarray}
g\int dx^{'}
\partial_{x} \phi(x,\tau)V(x-x^{'})\partial_{x^{'}} \phi(x^{'},\tau)
\end{eqnarray}
have the scaling dimension $x_{g}=\beta +1$ effectively,
which is consistent with the RG eq. of $g$ (2.10). 
The solution of eq. (2.10) is 
\begin{eqnarray}
g(l) &=& g(0) e^{(1-\beta)l}=g(0) e^{(1-\beta)\ln L}
=g(0)\frac{1}{L^{\beta-1}},
\end{eqnarray} 
where we use $l=\ln L$. Finally we obtain the accurate finite size scaling:
\begin{eqnarray}
\Delta E_{n} &=& \frac{2\pi v x_{n}}{L}(1+\frac{{\rm const.}}{L^{2(\beta-1)}}
+\frac{{\rm const.}}{L^{\beta+1}}+O(1/L^2))
\nonumber \\
E_{g} &=& e_{g}L -\frac{\pi v c}{6L}(1+\frac{{\rm const.}}{L^{2(\beta-1)}}
+\frac{{\rm const.}}{L^{\beta+1}}+O(1/L^2)).
\end{eqnarray} 
Moreover we can discuss the thermodynamics properties. We replace 
$ E_{g} \rightarrow f/T$ and $L\rightarrow v/T$ in 
the ground state energy of eq. (2.15), where $f$ is 
the free energy per temperature. We obtain the low temperature 
behaviors of $f$:
\begin{eqnarray}
f &=& -\frac{\pi v c}{6}\frac{T^{2}}{v^{2}}(1
+{\rm const.} (\frac{T}{v})^{2(\beta-1)}
+{\rm const.} (\frac{T}{v})^{(\beta+1)}
+{\rm const.} (\frac{T}{v})^2).
\end{eqnarray} 
Thus the specific heat $C=-T\frac{\partial^2 f}{\partial T^2}$ is
\begin{eqnarray}
C &=& \frac{\pi c T}{3 v}(1
+{\rm const.} (\frac{T}{v})^{2(\beta-1)}
+{\rm const.} (\frac{T}{v})^{(\beta+1)}
+{\rm const.} (\frac{T}{v})^2).
\end{eqnarray}
\section{Numerical calculations}
We have investigated the properties of the energy scaling and 
derived the corrections terms caused by the LR forward scatterings.
Let us consider 
the following tight-binding Hamiltonian with LR interactions:
\begin{eqnarray}
 H &=& -t \sum_{j} (c^{\dagger}_{j+1} c_{j} + {\rm h.c} )+
\frac{V}{2} \sum_{i\neq j} (n_{i}-<n>)
V(|i-j|) (n_{j}-<n>),
\end{eqnarray}
where $V(i-j)= \frac{1}{(\frac{L}{\pi} \sin \frac{\pi(i-j)}{L})^\beta}$
and $n_{j}=c^{\dagger}_{j}c_{j}$. And
we impose the periodic boundary condition and $<n>=1/2$.
By the straight forward bosonization technique, 
the effective action of (3.1) for the arbitrary filling can be written by
\begin{eqnarray}
 S &=& \int d\tau dx \frac{1}{2 \pi K} (\nabla \phi)^2 + g\int d\tau dx dx^{'}
\partial_{x} \phi(x,\tau)V(x-x^{'})\partial_{x^{'}} \phi(x^{'},\tau) 
\nonumber \\
   &+&g\;{\rm const.} \int d\tau dx dx^{'} 
\cos(2k_{F}x+{\sqrt 2}\phi(x,\tau))V(x-x^{'})
\cos(2k_{F}x^{'}+{\sqrt 2}\phi(x^{'},\tau)).
\end{eqnarray}
The last oscillating term 
consists of the umklapp process term 
$\cos {2 \sqrt 2} \phi$ and the longer range interactions. 
We have found the corrections in eqs. (2.15) caused by the
irrelevant LR forward scatterings ($\beta >1$ and $\beta \neq$ odd
integer) in the previous section.
However the oscillating terms may disturb the TLL and
cause the mass gap. 
We numerically investigate how long range the TLL phase survive for
the strength of the LR interactions.
By making use of eqs. (2.15), we can calculate the
compressibility $\chi=K/v$ and the drude weight $D=vK$ within
the TLL framework. 

The operators $\cos {\sqrt 2} \phi$ and $e^{\pm i \sqrt 2 \theta}$
have the scaling dimensions $K/2$ and $1/2K$ in 
the TLL. The operators 
have the symmetries $q=\pi,S_z=0$ and $q=\pi,S_z=1$ respectively. The 
explicit excitations describing the compressibility and the drude weight are
\begin{eqnarray}
 \chi &=& K/v=1/(2L\Delta E(S_z=1, q=\pi)) \nonumber \\
  D   &=& vK= 2 L\Delta E(q=\pi).
\end{eqnarray}
In Fig. 1 and 2 we plot the compressibility $\chi$ and the drude weight
$D=vK$ versus 
the interactions strength $g$ for the various powers $\beta$.
For $g < 0$ the $\chi$ exhibits the rapid increase which suggests the phase 
separation. In spin variables' language of (3.1), that is, XXZ model,
this phase separation is nothing but the ferromagnetic phase. 
Hence for the larger $\beta$ the 
point of the phase separation approaches to $-1$. 
For $g > 0$ we see the subtle tendency
that the $\chi$ become smaller as the $\beta$ is smaller.   
For the drude weight of $g > 0$ 
we find that the D become larger as the $\beta$ approach to the Coulomb 
interactions case $\beta=1$. 

In Fig. 3 we plot the velocity versus
the interactions strength $g$ for the various powers $\beta$.
The velocity is defined by
\begin{eqnarray}
 v&=&\frac{L}{2\pi}\Delta E(q=2\pi/L).
\end{eqnarray}
Note that the velocities are finite values for $\beta >1$ 
on the contrary of the Coulomb
interactions case $\beta=1$ (See APPENDIX B 2.). 
There are the points where the velocities 
are zero, implying the existences of the phase separation.

In Fig. 4 we plot the normalization $\frac{D}{\chi v^2}$ versus
the interactions strength $g$ for the various powers $\beta$, where
$D$, $\chi$ and $v$ are defined by eqs. (3.3) and eq. (3.4) respectively.
If the system belongs to TLL, this value is expected to be 1.
We see that the TLL region become wider as the $\beta$ 
approaches to $1$ for $g>0$.
Reversely for $g<0$ the TLL region is smaller as $\beta$ goes to $1$.
\section{Discussions and summary}
We have investigated the range of the TLL theoretically 
and numerically by utilizing the scaling argument of renormalization
group and CFT. We have found that the TLL region become wider 
and the drude weight become larger as the power 
$\beta$ approaches to 1 which is the Coulomb case. 
It was found experimentally\cite{Chandra}
that the amplitude of the persistent current in the micron-size 
Au loops is larger than a predicted value\cite{Cheung}.
Recently it has been found by the 
self harmonic approximation treatments that the localization length in
the sine-Gordon model with randomness become larger when the 
Coulomb interactions are considered\cite{Butusei}.
The our present results seem to give the additional confirmations 
to these features.

We have found that the TLL is broken for the stronger LR interactions.
We expect that the system is gapful for stronger $g$ 
and the ground state become two fold 
degenerate independently of $\beta (>1)$. 
Yamanaka {\it et al} derived the necessary conditions for
the gap generations 
in the fermion systems of one dimension with nonperturbative arguments
\cite{Yama}:
\begin{eqnarray}
 n\rho = \rm integer,
\end{eqnarray}
where $n$ is the period of ground state
and $\rho = N_{F}/L$ is the density of the fermion. 
Hence $n=2$ is derived for $\rho=1/2$ ($k_{F}=\pi/2$). 
This is reasonable, because the LR interactions 
include the short range umklapp process
term $\cos {2 \sqrt 2} \phi$ which cause the mass gap and 2 fold
degeneracy of the system. 

Let us argue the case $\beta =1$.
Assuming the g term is the small 
perturbation, we obtain the energy size scaling (See APPENDIX B and 
APPENDIX C.):
\begin{eqnarray}
\Delta E_{n} &=& \frac{2\pi v_{0} x_{n}}{L}(1+O(g)+g(0)\;{\rm const.}\ln(1/L)
)
\nonumber \\
E_{g} &=& e_{g}L -\frac{\pi v_{0} c}{6L}(1+O(g)+
g(0)\;{\rm const.}\ln(1/L)
),
\end{eqnarray}
where we assume $g(l) \sim g(0) \sim {\rm const.}$ (See APPENDIX B.).
We should interpret that the velocity $v_{0}$ is for no perturbations.
According to these results, 
the g terms break the TLL behaviors $L\Delta E= {\rm const.}$ for 
$L \rightarrow \infty$. This
estimation is consistent with a 
numerical report that TLL is broken by the Coulomb interactions\cite{Gia}. 

Through the replacements same as $\beta >1$ case,
we obtain the temperature dependences of the free energy and specific heat:
\begin{eqnarray}
f &=& -\frac{\pi c}{6 v_{0}}T^{2}(1+O(g)
+g\;{\rm const.} \ln(\frac{T}{v_{0}}))
\end{eqnarray}
and
\begin{eqnarray}
C &=& \{\frac{\pi c }{3 v_{0}}+O(g)\}T
+g \;{\rm const.} 
\frac{T}{v_{0}}\ln(\frac{T}{v_{0}}),
\end{eqnarray}
where const. is the positive constant (See the negative C term of eqs. (C9) 
in APPENDIX C.).
These suggest that the TLL is broken even 
for the slight $g$ when T is near to 0 
because $|g\frac{T}{v_{0}}\ln(\frac{T}{v_{0}})|$ become larger 
than $T$. 
When $T = 0$, even for small $g$, the system belongs to
other universality class in which the density correlation 
function is given by (2.5). 
However we can justify 
that the TLL holds if $O(e^{-1/g}) << T < 1$ 
\footnote{Though the scaling (C5) is valid for finite size
(finite temperature), we are treating the lower temperature behaviors
now. Therefore we should write $O(e^{-1/g})<< T < O(1)$. 
}
because the $|g\frac{T}{v_{0}} \ln(\frac{T}{v_{0}})|$
is smaller than $T$. For this case it is convincing that 
the 
$g \frac{T}{v_{0}}\ln(\frac{T}{v_{0}})$ terms 
are the small perturbations. Therefore we can have the pictures 
for the case $\beta =1$ (See Fig. 8.). 

In summary, by utilizing the bosonization 
technique, the renormalization group and the CFT
we analyzed the TLL with LR forward scatterings. And 
we could obtain the accurate finite size scaling of the energies 
and thermodynamics properties.
By making use of these scaling relations and the numerical 
calculations we found that the range of TLL and the drude weight
increase as the interactions' power $\beta$ approaches to 1 which 
is the Coulomb interactions' one. Furthermore we obtained the 
specific heat for $\beta =1$, which 
deviates from the linear $T$ to T+g$T\ln T$, where the small $g$ takes 
the positive value if the LR interactions are repulsive 
. This implys the TLL holds when $O(e^{-1/g}) << T < 1$.
\acknowledgements
I thank K. Nomura for his encourages. The numerical 
calculations in this work 
is based on the package TITPACK ver. 2.0 by H. Nishimori. 
\appendix
\widetext
\section{The long wavelength behavior of $V(q)$}
The Fourier form of the interaction $V(x)$ is given by 
\begin{eqnarray}
 V(q) &=& 2 \frac{\sqrt{\pi} \alpha^{\beta/2-1/2}}{\Gamma(\beta/2) 
2^{\beta/2-1/2}} 
q^{\beta/2-1/2} K_{\beta/2-1/2}(\alpha q),
\end{eqnarray}
where $K_{\nu}(x)$ is the modified Bessel function of $\nu$th order
and $\Gamma(x)$ is the gamma function.
When $\beta=1$, $V(q)=K_{0}(\alpha q)\sim - \ln q\;\; (q \rightarrow 0)$
which was already discussed by Schulz \cite{Schulz}.
When $\beta > 1$ and $\beta \neq $ odd integer,
\widetext
\begin{eqnarray}
V(q) &=& \int_{-\infty}^{\infty} dx \frac{e^{iqx}}{(x^2+\alpha^2)^{\beta/2}}=\int dx \frac{1}{(x^2+\alpha^2)^{\beta/2}}+\int dx 
\frac{e^{iqx}-1}{(x^2+\alpha^2)^{\beta/2}} \nonumber \\
&=& {\rm const.}+\frac{1}{q^{1-\beta}}\int dx^{'}
\frac{e^{i x^{'}}-1}{  (x^{'2}+ q^2 \alpha^2)^{\beta/2}    } 
\\ \nonumber
&\sim& {\rm const.} + A q^{\beta-1},\;A:{\rm const.}
\end{eqnarray}
When $\beta$ is odd integer, the $V(q)$ is shown in the main part of the 
present paper. 
\section{Renormalization group equation}
\subsection{$\beta > 1$ and $\beta \neq$ odd integer}
We derive the renormalization group equations heuristically.
Let us start from the action (2.9): 
\begin{eqnarray}
 S &=& \sum_{w} \sum_{q=-\Lambda}^{\Lambda} \frac{2\pi}{K^{'}}
(q^2+w^2)|\phi(q,w)|^2+g\sum_{w} \sum_{q=-\Lambda}^{\Lambda}q^2 
(V(q)-V_{\rm short})|\phi(q,w)|^2 \nonumber \\ 
   &=& \sum_{w}\{\sum_{q=-\Lambda/b}^{\Lambda/b}
+\sum_{q=-\Lambda}^{-\Lambda/b}+\sum_{q=\Lambda/b}^{\Lambda}\}+g
\sum_{w}\{\sum_{q=-\Lambda/b}^{\Lambda/b}
+\sum_{q=-\Lambda}^{-\Lambda/b}+\sum_{q=\Lambda/b}^{\Lambda}\}.
\end{eqnarray}
The partition function is
\begin{eqnarray}
Z &=& \int {\cal D} \phi_{slow} {\cal D}\phi_{fast}\exp (-
S^{0}_{slow}
-S^{0}_{fast}-S^{g}_{slow}-S^{g}_{fast}). 
\end{eqnarray}
Thus we can integrate out $S_{fast}$ 
($|q|>\Lambda/b$ component) simply and obtain 
\begin{eqnarray}
Z &=& \int {\cal D} \phi_{slow} \exp (-S^{0}_{slow}-S^{g}_{slow}).
\end{eqnarray} 
The remaining procedure of the renormalization is the scale
transformation
\begin{eqnarray}
q \rightarrow q/b, \;w\rightarrow w/b \;\;{\rm and}\;\;
\phi \rightarrow \phi b^2, 
\end{eqnarray}
where we choose the dynamical exponent 1. The results are 
\begin{eqnarray}
S^{0}_{slow} &\rightarrow & S^{0}\nonumber \\
S^{g}_{slow} &\rightarrow & g\sum_{w}\sum_{q=-\Lambda}^{-\Lambda}
q^2(V(q/b)-V_{\rm short})|\phi(q,w)|^2 \nonumber \\
&\rightarrow & gb^{1-\beta}\sum_{w}\sum_{q=-\Lambda}^{-\Lambda}
q^2(V(q)-V_{\rm short})|\phi(q,w)|^2,
\end{eqnarray}
where we use the behavior $V(q)-V_{\rm short}\sim q^{\beta-1}$ 
from (2.4). Hence we obtain the renormalization group eq.
\begin{eqnarray}
\frac{dg(b)}{d b} = (1-\beta)\frac{g(b)}{b}.
\end{eqnarray}
We put $l=\ln b$ and obtain eq. (2. 10).
\subsection{$\beta =1$}
The dispersion relations of 
the Coulomb interactions case include the marginal part $w \sim q$ and 
$w \sim q \sqrt{\ln q}$ as well as the $\beta >1$ case. However 
it is difficult to know the explicit separated function like 
eqs. (2.7) and (2.8). Therefore we renormalize by using the bare 
$V(q) \sim -A \log q + B$.

Integrating out the fast part, we obtain the effective action of the
slow part
\begin{eqnarray}
 S_{\rm slow} &=& \sum_{w} \sum_{q=-\Lambda/b}^{\Lambda/b} \frac{2\pi}{K}
(v q^2+w^2/v)|\phi(q,w)|^2+g\sum_{w} \sum_{q=-\Lambda/b}^{\Lambda/b}q^2
V(q)|\phi(q,w)|^2, \nonumber \\
\end{eqnarray}
where we dare to leave the velocity in the Gaussian part. Note that
we need not the renormalization of the velocity in the case $\beta >1$.
After the scale transformation (B 4), we obtain the eqs.
\begin{eqnarray}
 \frac{dg}{dl}&=&0 \nonumber \\
 \frac{d}{dl}(\frac{v}{K})&=&\frac{gA}{2\pi} \nonumber \\
 \frac{d}{dl}(\frac{1}{vK})&=&0.
\end{eqnarray}
We see that the $K$ and the velocity $v$
is renormalized in stead of the no renormalization of g.
The velocity v can be written by
\begin{eqnarray}
v(b) &=& \sqrt{ {\rm const.}+g\; {\rm const.} \ln b} \sim \sqrt{\ln L}.
\end{eqnarray}
The velocity shows the weak divergence for long distances, 
which is consistent with the estimations of $v=\frac{dw}{dq}$ from
the eqs. (2.3) and (2.4).
\section{The finite size scaling of energy}
We write the Hamiltonian in the finite strip from the action (2.1): 
\begin{eqnarray}
H &=& H_{\rm TLL}
+g\int_{-L/2}^{L/2}d\sigma_{1}d\sigma_{2}\partial_{\sigma_{1}}\phi(\sigma_{1})
\partial_{\sigma_{2}}\phi(\sigma_{2})
V(|\sigma_{1}-\sigma_{2}|)\theta(|\sigma_{1}-\sigma_{2}|-\alpha_{0}),
\end{eqnarray}
where the $H_{\rm TLL}$ is the TLL part of the Hamiltonian. We introduce
the step function $\theta(x)$ to avoid the ultra violet divergences
which come from $V(x)$ and the operator 
product expansion of $\partial_{\sigma}\phi(\sigma)$.
For the small perturbation $g$ the ground state energy
$E_{g}$ varies as
\begin{eqnarray}
  E^{'}_{g}- E_{g}&=& 
g\int_{-L/2}^{L/2}d\sigma_{1}d\sigma_{2}V(|\sigma_{1}-\sigma_{2}|)
<0|\partial_{\sigma_{1}}\phi(\sigma_{1})\partial_{\sigma_{2}}\phi(\sigma_{2})|0> \theta(|\sigma_{1}-\sigma_{2}|-\alpha_{0})
\nonumber \\
  &=&g\int_{-L/2}^{L/2}d\sigma_{1}d\sigma_{2}V(|\sigma_{1}-\sigma_{2}|)
[ <0|\partial_{w_{1}}\varphi(w_{1})\partial_{w_{2}}\varphi(w_{2})|0> 
\nonumber \\
  & & + <0|\partial_{\bar{w}_{1}}\bar{\varphi}(\bar{w}_{1})
\partial_{\bar{w}_{2}} \bar{\varphi}(\bar{w}_{2})|0> ]_{\tau_{1}=\tau_{2}=0}
\theta(|\sigma_{1}-\sigma_{2}|-\alpha_{0}),
\end{eqnarray}
where we introduce the coordination $w=\tau+i\sigma$ (
$-L/2< \sigma <L/2$, $-\infty < \tau < \infty$). 
From the characters of the Gaussian part (TLL part) we can 
separate as $\phi(\sigma,\tau)=
\varphi(w)+\bar{\varphi}(\bar{w})$ and derive
$<0|\partial_{\bar{w}_{1}}\bar{\varphi}(\bar{w}_{1})
\partial_{w_{2}} \varphi(w_{2})|0>=0$.
The content of the brackets is modified as follows:
\widetext
\begin{eqnarray}
&&[ <0|\partial_{w_{1}}\varphi(w_{1})\partial_{w_{2}}\varphi(w_{2})|0>+
<0|\partial_{\bar{w}_{1}}\bar{\varphi}(\bar{w}_{1})
\partial_{\bar{w}_{2}} \bar{\varphi}(\bar{w}_{2})|0> 
]_{\tau_{1}=\tau_{2}=0} \nonumber \\
&=& [  (\frac{2\pi}{L})^{2\Delta} \frac{z_{2}}{z_{1}}
\frac{1}{(1-\frac{z_{2}}{z_{1}})^2} 
+ (\frac{2\pi}{L})^{2\bar{\Delta}} \frac{\bar{z}_{2}}{\bar{z}_{1}}
\frac{1}{(1-\frac{\bar{z}_{2}}{\bar{z}_{1}})^2}  
]_{\tau_{1}=\tau_{2}=0} \nonumber \\
&=& -(\frac{2\pi}{L})^{2}\frac{1}{2} \frac{1}{\sin^2 \frac{\pi 
(\sigma_{1}-\sigma_{2})}{L} },
\end{eqnarray}
where we transform the correlations
$<\partial_{z_{1}}\varphi(z_{1})\partial_{z_{2}}\varphi(z_{2})>=1/(z_{1}-z_{2})^2$
in $\infty \times \infty$ $z$ plane
to the present strip $w$ thorough $z=\exp{\frac{2\pi w}{L}}$. 
At present case $\partial_{w}\varphi(w)$
($\partial_{\bar{w}}\bar{\varphi}(\bar{w})$)
have the spin $s=1(-1)$ and conformal dimension $\Delta=1$($\bar{\Delta}=1$). 
Hence we obtain
\begin{eqnarray}
  E^{'}_{g}-E_{g}&=& 
\frac{g}{8} (\frac{2\pi}{L})^{2} L 
\int_{-L}^{L}d x
V(|x|)\frac{1}{\sin^2 \frac{\pi x}{L}}
\theta(|x|-\alpha_{0}) \nonumber \\
&=& \frac{g}{8} (2\pi)^{2}\int_{-1}^{1}dx^{'}V(|Lx^{'}|)
\frac{1}{\sin^2 \pi x^{'}}\theta(L|x^{'}|-\alpha_{0}) \nonumber \\
&=&
\frac{g}{8} (2\pi)^{2}  (\frac{\pi}{L})^\beta  \int_{-1}^{1}dx^{'}
\frac{1}{(\sin \pi |x^{'}|)^{\beta}}
\frac{1}{\sin^2 \pi x^{'}}\theta(|x^{'}|-\frac{\alpha_{0}}{L}),
\end{eqnarray}
where we impose the periodic boundary 
condition and use the interaction potential $V(x)=
1/(\frac{L}{\pi}\sin(\frac{x \pi}{L}))^\beta$. 
Putting $\epsilon=\alpha_{0}/L$ for convenience, we give 
the differential of the integral part: 
\begin{eqnarray}
&& \frac{\partial}{\partial \epsilon}
\int_{-1}^{1}dx^{'}
\frac{1}{(\sin \pi |x^{'}|)^{\beta}}
\frac{1}{\sin^2 \pi x^{'}}\theta(|x^{'}|-\epsilon) \nonumber \\
&=&-\frac{1}{(\sin \pi |\epsilon|)^{\beta}}
\frac{1}{\sin^2 \pi \epsilon}
\end{eqnarray}
After integrating the Taylor expansion about $\epsilon$ of this
quantity, we obtain
\begin{eqnarray}
&& \int_{-1}^{1}dx^{'}
\frac{1}{(\sin \pi |x^{'}|)^{\beta}}
\frac{1}{\sin^2 \pi x^{'}}\theta(|x^{'}|-\epsilon) \nonumber \\
&=&\frac{1}{\pi}[{\rm const.}-\frac{(\pi \epsilon)^{-\beta-1}}{-\beta-1}
-\frac{\beta+2}{3!(-\beta+1)}  (\pi \epsilon)^{-\beta+1} \nonumber \\
&&
+\frac{1}{-\beta+3} \{ -\frac{1}{5!}(\beta+2)+\frac{1}{72}(\beta++1)(\beta+2)\}
(\pi \epsilon)^{-\beta+3}+O((\pi \epsilon)^{-\beta+5})
],
\end{eqnarray}
where $\beta \neq$ odd integer.
Therefore we can write the corrections in the form:
\begin{eqnarray}
E^{'}_{g} - E_{g}&=& g[ \frac{{\rm A}}{L^\beta}
+{\rm B}L
+{\rm C}\frac{1}{L}+D\frac{1}{L^3}+O(\frac{1}{L^5})],
\end{eqnarray}
where A, B, C and D are the finite constant values. 
The form of B and C terms is just same as the second term 
of eqs. (2.11). We assume that these terms should be renormalized to
TLL, because the LR interactions 
inevitably contain the short range types of interactions 
which are reduced to $(\partial_{x}\phi)^2$. It is natural and
reasonable that in the corrections of the ground state energy
there are the same contributions as the TLL forms of (2.11).  
Thus we think that the intrinsic contributions of the LR interactions are 
\begin{eqnarray}
E^{'}_{g} - E_{g}&=& 
g[ \frac{{\rm A}}{L^\beta}+B\frac{1}{L^3}+O(\frac{1}{L^5})].
\end{eqnarray}
This result is consistent with the spectrum analysis
and the RG results for $\beta >1$ \footnote{If the B and C term 
in eq. (C. 7) 
give the important contributions to
the excitations, 
the TLL is broken by such the contributions. This
is discrepant with the spectrum results and the RG arguments.
Hence we assume that the B and C contributions
in eq. (C. 7) are not intrinsic. Especially we should 
regard the B term nonuniversal bulk constants}.
For the $\beta$= odd number case, there exist the logarithmic
corrections instead of the eq. (C. 6). 
We write the results for the respective $\beta$ specifically:
\begin{eqnarray}
E^{'}_{g} - E_{g}&=&
g[ \frac{{\rm A}}{L}+{\rm B}L
+\frac{{\rm C}}{L}\ln \frac{1}{L}+D\frac{1}{L^3}+O(\frac{1}{L^5})]\;\; 
{\rm for\;\;\beta=1} \nonumber \\
&&g[ \frac{{\rm A}}{L^3}+{\rm B}L
+\frac{{\rm C}}{L}
+D\frac{1}{L^3}\ln\frac{1}{L}
+O(\frac{1}{L^5})]\;\;{\rm for\;\;\beta=3} \nonumber \\
&&g[ \frac{{\rm A}}{L^5}+{\rm B}L
+\frac{{\rm C}}{L}
+D\frac{1}{L^3}+E\frac{1}{L^5}\ln\frac{1}{L}+O(\frac{1}{L^7})]\;\;
{\rm for\;\;\beta=5} \nonumber \\
&& \cdots
\end{eqnarray}   

Next we derive the corrections for the energy of the excited state:
\begin{eqnarray}
E_{n}^{'}-E_{n}&=&
g
\int_{-L/2}^{L/2}d\sigma_{1}d\sigma_{2}V(|\sigma_{1}-\sigma_{2}|)
<n|\partial_{\sigma_{1}}\phi(\sigma_{1})\partial_{\sigma_{2}}\phi(\sigma_{2})|n>\theta(|\sigma_{1}-\sigma_{2}|-\alpha_{0})
\nonumber \\
&=& g
\int_{-L/2}^{L/2}d\sigma_{1}d\sigma_{2}V(|\sigma_{1}-\sigma_{2}|)
\sum_{\alpha}
<n|\partial_{\sigma_{1}}\phi(\sigma_{1})|\alpha><\alpha|
\partial_{\sigma_{2}}\phi(\sigma_{2})|n>
\theta(|\sigma_{1}-\sigma_{2}|-\alpha_{0})
\nonumber \\
&=& g\sum_{\alpha}C_{n j \alpha} C_{\alpha j n} (\frac{2\pi}{L})^{2}
\int_{-L/2}^{L/2}d\sigma_{1}d\sigma_{2}
V(|\sigma_{1}-\sigma_{2}|)
e^{2\pi i(s_{n}-s_{\alpha})(\sigma_{1}-\sigma_{2})/L}
\theta(|\sigma_{1}-\sigma_{2}|-\alpha_{0})
\nonumber \\
&=& g \sum_{\alpha}C_{n j \alpha} C_{\alpha j n}
(\frac{2\pi}{L})^{2} L\int_{0}^{L}dx V(|x|)
\cos \frac{2\pi}{L}(s_{n}-s_{\alpha})x
\;\;\theta(|x|-\alpha_{0}) \nonumber \\
&=& g \sum_{\alpha}C_{n j \alpha} C_{\alpha j n} \frac{(2\pi)^2}{L^{\beta}}
\int_{0}^{1}dy \frac{1}{(\sin \pi |y|)^\beta}
\cos 2\pi(s_{n}-s_{\alpha})y\;\;\theta(|y|
-\frac{\alpha_{0}}{L})\nonumber \\
&=& g \sum_{\alpha}C_{n j \alpha} C_{\alpha j n}
\{ \frac{A(s_{n}-s_{\alpha})}{L^\beta}+\frac{B}{-\beta+1}\frac{1}{L} 
+\frac{C(s_{n}-s_{\alpha})}{-\beta+3}\frac{1}{L^3}
\nonumber \\
&&+\frac{D(s_{n}-s_{\alpha})}{-\beta+5}\frac{1}{L^5}+O(\frac{1}{L^7})\}
\;\; \beta \neq {\rm odd \;\;integer } \nonumber \\
&& g \sum_{\alpha}C_{n j \alpha} C_{\alpha j n}
\{ \frac{A(s_{n}-s_{\alpha})}{L}+B\frac{1}{L}\ln \frac{1}{L}
+C(s_{n}-s_{\alpha})\frac{1}{L^3} \nonumber \\
&&+D(s_{n}-s_{\alpha})\frac{1}{L^5}+O(\frac{1}{L^7})\}\;\;\beta=1  \nonumber \\
&& g \sum_{\alpha}C_{n j \alpha} C_{\alpha j n}
\{ \frac{A(s_{n}-s_{\alpha})}{L^3}+B\frac{1}{L}
+C(s_{n}-s_{\alpha})\frac{1}{L^3} \ln \frac{1}{L} \nonumber \\
&&+D(s_{n}-s_{\alpha})\frac{1}{L^5}+O(\frac{1}{L^7})\}\;\;\beta=3  \nonumber \\
&& g \sum_{\alpha}C_{n j \alpha} C_{\alpha j n}
\{ \frac{A(s_{n}-s_{\alpha})}{L^5}+B\frac{1}{L}
+C(s_{n}-s_{\alpha})\frac{1}{L^3} 
\nonumber \\
&&+D(s_{n}-s_{\alpha})\frac{1}{L^5}\ln \frac{1}{L} 
+O(\frac{1}{L^7})\}\;\;\beta=5  \nonumber \\
&&\cdots ,
\end{eqnarray}
where we use the results by Cardy\cite{Cardy4}:
\begin{eqnarray}
<n|\phi(\sigma)|\alpha>&=&C_{nj\alpha}(\frac{2\pi}{L})^{x_{j}}e^
{\frac{2\pi i(s_{n}-s_{\alpha})\sigma}{L}}.
\end{eqnarray}
We find the dependences of $1/L$ which should be renormalized 
to TLL. We think that this is caused by  
the LR interactions which include the short range interactions. 
The short interactions is nothing but the part of TLL. Hence
it is plausible for the $1/L$ dependences to appear.
These situations are same as the ground state properties
that we discussed just previously. Strictly speaking,
in the eqs. (C10) there is an ambiguity 
whether the $\sum_{\alpha}$ commute with 
the integrals. However it is difficult to prove 
the communtations unexpectedly, because we must 
know whether ths sum of 
the operator product expansion coefficients $\sum C_{nj\alpha}^2$ 
is finite, or 
not. We would transfer this problems to future works.  


\begin{figure}
\psfig{file=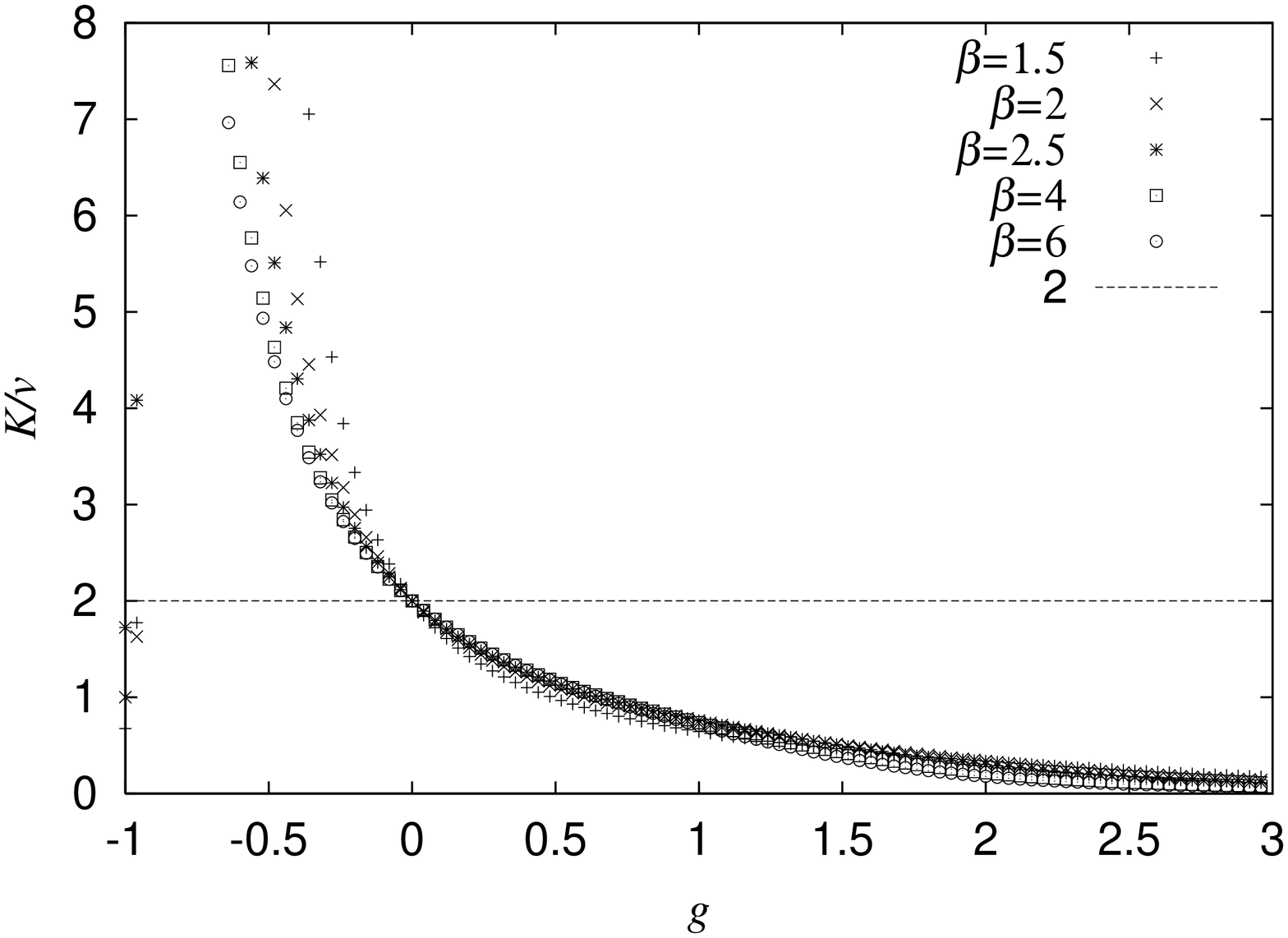,height=10cm,width=12cm,angle=-90,clip=}
\caption{
The extrapolated compressiblilty $\chi = K/v$ 
is plotted versus the strength $g$.
We use the size dependence $K/v(L)=K/v(\infty)
+\frac{a}{L}+\frac{b}{L^{2}}$ for $\beta =1.5$,
where $a$ and $b$ are determined
numerically (See eqs. (2.15).). 
We use  
$K/v(L)=K/v(\infty)
+\frac{a}{L^{2}}+\frac{b}{L^{3}}$ for $\beta =2, 2.5$ 
and $K/v(L)=K/v(\infty)
+\frac{a}{L^{2}}+\frac{b}{L^{4}}$ for $\beta \geq 4$. }
\end{figure}
\begin{figure}
\psfig{file=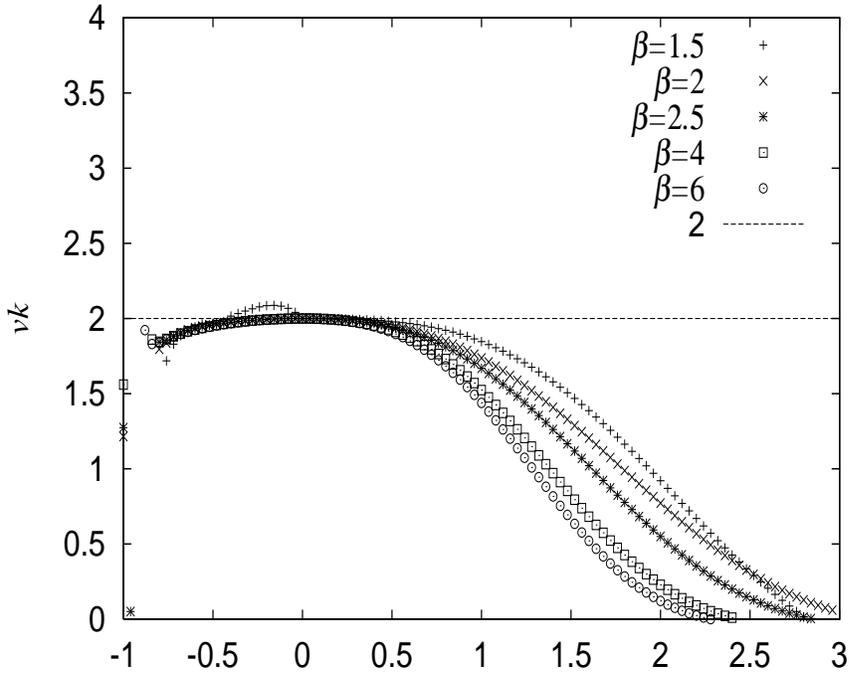,height=10cm,width=12cm,angle=-90,clip=}
\caption{The extrapolated drude weight $D=vK$ is plotted versus the strength
 $g$. We use the same scaling as compressiblilty.}
\end{figure}
\begin{figure}
\psfig{file=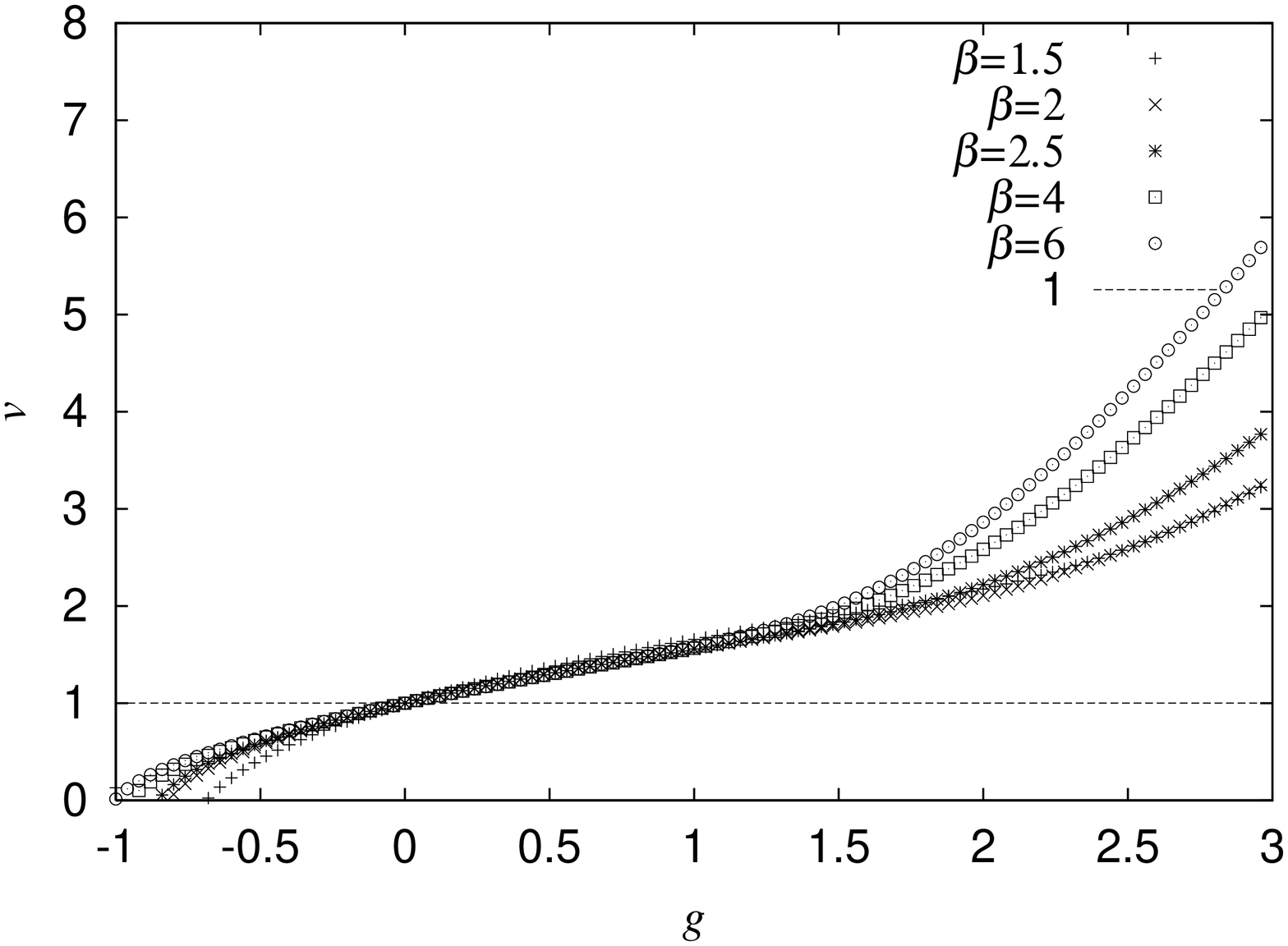,height=10cm,width=12cm,angle=-90,clip=}
\caption{The spin wave velocity $v$ is plotted versus the strength $g$.
We use same scaling as compressiblilty.}
\end{figure}
\begin{figure}
\psfig{file=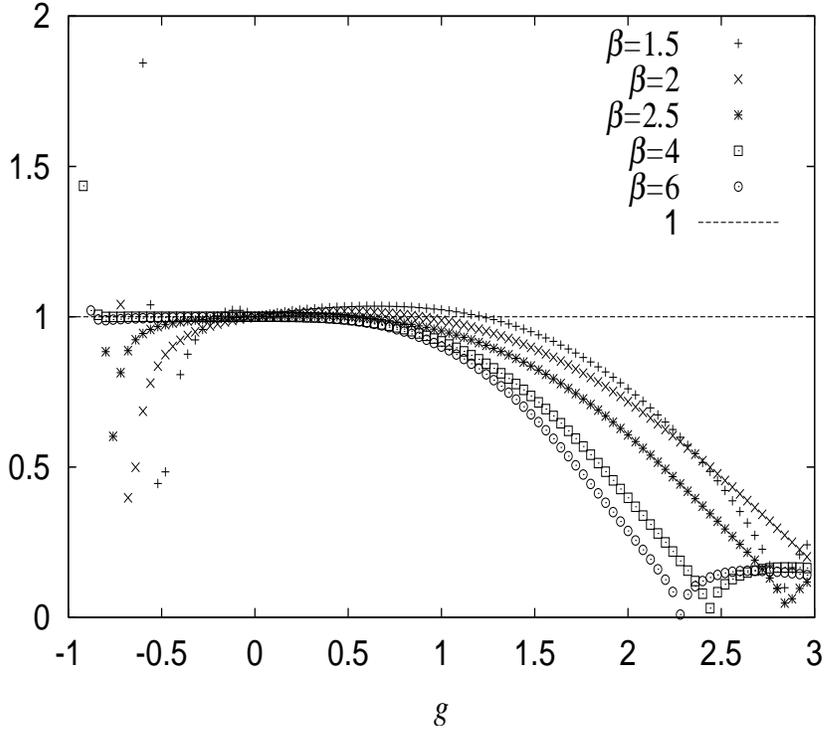,height=10cm,width=12cm,angle=-90,clip=}
\caption{The normalization $ \frac{D}{\chi v^2}$ is plotted versus the
 strength $g$.}
\end{figure}
\begin{figure}
\psfig{file=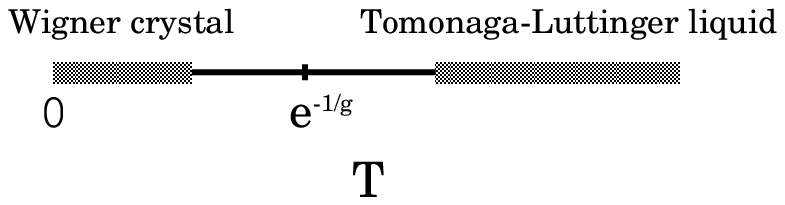,height=9cm,width=13cm,angle=0,clip=}
\caption{The schematic phase diagram versus temperature when 
$\beta =1$. Though the TLL generally lies for the low temperature $T$, 
the TLL is broken as $T \rightarrow 0$ when $\beta=1$. Instead 
the {\it Wigner crystal} phase emerges for the lower temperature
side than $O(e^{-1/g})$. However the TLL holds when $O(e^{-1/g})<<T<1$.
The value $O(e^{-1/g})$ is not the transition point but the point
where the perturabation theory breaks down.}
\end{figure}
\end{document}